\documentclass[reprint, superscriptaddress, tterverbose, amsmath,amssymb, aps, pra, floatfix,linenumbers,]{revtex4-2}

\usepackage{graphicx} 
\usepackage{physics} 
\usepackage{amsmath} 
\usepackage{dcolumn}
\usepackage{bm}
\usepackage{lipsum}

\usepackage{float}
\usepackage{soul} 
\usepackage{xcolor}

\begin{document}
\nolinenumbers 
\preprint{APS/123-QED}

\title{Phase and coherence retrieval from near- and far-field intensities}

\author{Eran Bernstein}
\email{Email: eran.bernstein@gmail.com} 
\author{Amit Pando}
\author{Asher A. Friesem}
\author{Nir Davidson}
\affiliation{Department of Physics of Complex Systems, Weizmann Institute of Science, Rehovot 761001, Israel}
\date{\today}

\begin{abstract}
Quantifying the coherence properties of complex optical fields is essential for applications ranging from high-power laser arrays to quantum coherent systems. Here, we present a new paradigm for coherence retrieval inspired by the Gerchberg–Saxton (GS) framework, enabling reconstruction of the first-order spatial coherence function (mutual intensity) of partially coherent light from intensity measurements in the near- and far-fields.
We introduce two complementary approaches: a four-dimensional Tensor GS algorithm that directly reconstructs the mutual intensity with high accuracy, and a Monte Carlo GS variant that significantly reduces computational cost at the expense of controlled approximation.
We validated both methods by reconstructing partially coherent fields in simulated linear and ring arrays of up to N=600 beams with prescribed Gaussian-decaying coherence. Experimentally, we applied the Tensor GS method to a triangular array of 130 coupled lasers with inhomogeneous spatial coherence, achieving good agreement with theory and a root-mean-square phase error as low as $2\pi/250$.
\end{abstract}

\maketitle

\section{Introduction}
Phase retrieval from intensity-only measurements is a fundamental problem in optics and wave physics, with applications in astronomy \cite{Dainty_1987}, ultrafast pulse characterization \cite{raz_2011}, laser physics \cite{Tradonsky_2019}, imaging through turbid media \cite{Bertolotti_2012,Katz_2014}, and Bose–Einstein condensates (BECs) \cite{Damm_2017}. Since the seminal work of Gerchberg and Saxton (GS) \cite{Gerchberg_1972a} and subsequent refinements by Fienup \cite{Fienup_1978, Fienup_1982}, iterative phase retrieval algorithms have become standard tools for reconstructing complex fields from near- and far-field intensity measurements \cite{Wang_1990,Ivanov_1992, Levi_1984, Bauschke_2002,Marchesini_2007}.

Most phase retrieval techniques assume fully coherent fields. But many important systems — including large laser arrays \cite{Nixon_2013}, X-ray and electron beams \cite{Wolf_2009,Nugent_2010,Latychevskaia_2017,Chen_2020,Xu_2024}, and structured quantum states \cite{Thibault_2013} — exhibit only partial spatial coherence. In such cases, a two-dimensional field must be described by a four-dimensional mutual intensity \( J(\mathbf{r},\mathbf{r}') = \langle E^*(\mathbf{r}) E(\mathbf{r}') \rangle \), i.e. the first-order spatial coherence function (or equivalently, the cross-spectral density in the frequency domain) \cite{Wolf_1982, Wolf_2008}. Existing reconstruction approaches, however, typically recover only the phase distribution  of a single coherent mode, rather than the full mutual intensity.

Several methods address partially coherent fields under restrictive assumptions. Mixed-state ptychography reconstructs low-rank density matrices using scanning redundancy and tomographically complete measurements \cite{Thibault_2013,Xu_2024}. Modal decomposition techniques can determine coefficients in a prescribed orthogonal basis from near- and far-field data, relying on strong prior knowledge of the coherence \cite{Cutolo_1995}. Extensions of the transport-of-intensity equation (TIE) to partially coherent illumination recover only phase information under restrictive prior assumptions \cite{Zuo_2015}. To our knowledge, there is no general method for reconstructing the full spatial mutual intensity from intensity measurements in only two planes.

Here, we present an approach for reconstructing the full mutual intensity of partially coherent fields from intensity measurements in two Fourier-conjugate planes (near-field and far-field). It is based on iterative propagation-and-projection of the mutual intensity between two planes and is applicable to spatially inhomogeneous partially coherent fields.
We develop two complementary implementations. The first, termed Tensor GS, directly reconstructs the mutual intensity with high accuracy. The second, termed Monte Carlo (MC) GS, is inspired by stochastic wave-function approaches to density-matrix reconstruction \cite{Molmer_1993} and significantly reduces computational complexity while providing accurate and controlled approximations.

We validated both methods using simulated linear and ring arrays of up to \(N=600\) beams with prescribed Gaussian-decaying coherence. Finally, we demonstrated experimental reconstruction in a triangular array of 130 coupled lasers, achieving excellent agreement with theory and a root-mean-square phase error as low as \(2\pi/250\). These results indicate that our approach is practical for characterizing large-scale partially coherent optical systems.

\section{Methods}

Our approach for reconstructing the spatial mutual coherence function of a partially coherent field $E(\mathbf{r})$ from intensity measurements is based on the GS algorithm adapted for the cross-spectral density of the field $W(\mathbf{r},\mathbf{r'},\omega)$. We consider the case of quasi-monochromatic fields where the cross-spectral density reduces to the mutual intensity $J(\mathbf{r},\mathbf{r'})$, defined as 

\begin{equation}
    \begin{split}
        & W(\mathbf{r},\mathbf{r'},\omega) = <E^*(\mathbf{r},\omega)E(\mathbf{r'},\omega)>,\\
        & J(\mathbf{r},\mathbf{r'}) = <E^*(\mathbf{r})E(\mathbf{r'})>,
    \end{split}
    \label{Eq0_crossDensity2Coherence}
\end{equation}

where brackets denote the ensemble average, or equivalently, the time-averaged measurement. The diagonal elements of the near-field mutual intensity correspond to the near-field intensity ($I^{NF}(\mathbf{r})=J(\mathbf{r},\mathbf{r})$) and its Fourier transform is related to the far-field mutual intensity, up to a minus sign in one of the momentum coordinates, ($\tilde{J}(\mathbf{k},\mathbf{k'})=\text{FT}\{J(\mathbf{r},\mathbf{r'})\}(-\mathbf{k},\mathbf{k'})$) \cite{Wolf_2008}. Our basic approach is to iteratively propagate the full four-dimensional mutual intensity between near- and far-field planes while projecting the diagonals on the measured (target) intensities in both planes. From the mutual intensity, we obtain the normalized first-order spatial mutual coherence as

\begin{equation}
    \begin{split}
        & g_1(\mathbf{r},\mathbf{r'}) = \frac{J(\mathbf{r},\mathbf{r'})}{\sqrt{J(\mathbf{r},\mathbf{r}) J(\mathbf{r'},\mathbf{r'})}}.
    \end{split}
    \label{Eq1_crossDensity2Coherence}
\end{equation}

\begin{figure}[ht!]
    \centering
    \includegraphics[width=1\linewidth]{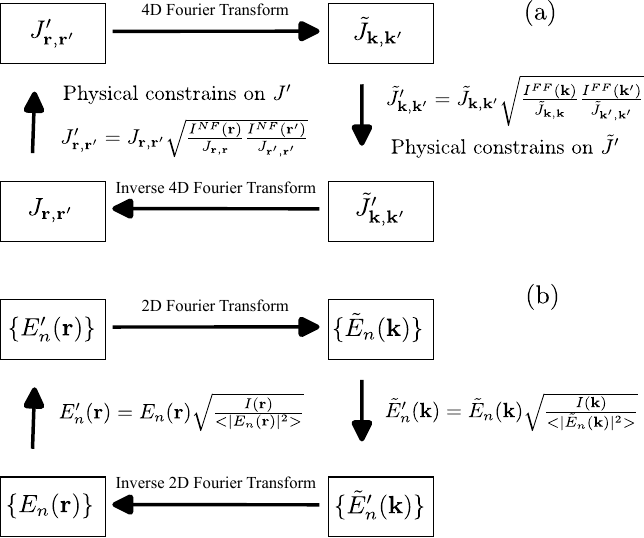}
    \caption{Iterative reconstruction steps for (a) Tensor GS and (b) MC GS. $I^{NF}(\mathbf{r})$ and $I^{FF}(\mathbf{k})$ are the target near-field and far-field intensities. $J_{\mathbf{r},\mathbf{r'}}$ and $J_{\mathbf{k},\mathbf{k'}}$ denote $J(\mathbf{r},\mathbf{r'})$ and $J(\mathbf{k},\mathbf{k'})$ respectively.}
    \label{fig:1_TensorGS_Algo}
\end{figure}

We develop two methods for reconstructing both phase and coherence: the Tensor GS method and the Monte Carlo (MC) GS variant. The Tensor GS method applies GS-like alternating projections directly to the mutual intensity, which is iteratively propagated between the near-field and far-field planes.
Starting from an initial mutual intensity with vanishing off-diagonal elements, it is propagated to the far-field via a four-dimensional Fourier transform. The target far-field intensity is then enforced by projecting onto the diagonal elements of the mutual intensity. Simultaneously, the off-diagonal elements are updated to preserve the underlying coherence structure during the projection step, using the relations shown in Fig.~\ref{fig:1_TensorGS_Algo}(a).
Next, Hermitian symmetry is imposed by symmetrizing the mutual intensity and the off-diagonal elements are constrained to ensure physical consistency of the coherence function, enforcing $|\tilde{g}_1(\mathbf{k},\mathbf{k'})|\leq 1$. The updated far-field mutual intensity is then inverse Fourier transformed back to the near-field, where an analogous projection step is applied. This iterative process is repeated until convergence.

To reduce the computational complexity associated with four-dimensional Fourier transforms, we introduce a Monte Carlo (MC) GS method. In this approach, the partially coherent field is represented by an ensemble of $P$ independent random complex realizations, initialized without prior information about the target field, in analogy to the well-established stochastic  wave-function approach of \cite{Molmer_1993}. 
At each iteration, all fields in the ensemble are propagated between the near-field and far-field planes using standard two-dimensional Fourier and inverse Fourier transforms. After each propagation step, the field amplitudes are multiplied by a common function such that the ensemble-averaged intensity matches the target intensity in the corresponding plane, as illustrated in Fig.~\ref{fig:1_TensorGS_Algo}(b). 
The mutual intensity defined in Eq.~(\ref{Eq0_crossDensity2Coherence}) is then estimated by ensemble averaging over $P$ realizations. The accuracy of the reconstructed coherence therefore scales as $\frac{1}{\sqrt{P}}$, enabling systematic control of the approximation by increasing the ensemble size, while maintaining reduced computational complexity. For a field composed of $M$ spatial modes, the MC GS method scales as $PM\log(M)$, compared to the $M^2\log(M)$ scaling of the Tensor GS method.

The performance of our reconstruction methods is compared against the classical inverse Fourier transform (IFT) of the far-field intensity, which accurately recovers the coherence function only under the assumptions of spatially homogeneous coherence and a uniform near-field intensity distribution \cite{Wolf_2013}. For inhomogeneous coherence, the IFT approach fails to reconstruct the correct mutual intensity while our methods do not fail. For partial coherence that arises from a small number of fully coherent modes, a relatively simple and intuitive reconstruction procedure is possible, as described in Section 1 of Supplement.

\section{Numerical Results}

\begin{figure}[ht!]
    \centering
    \includegraphics[width=0.6\linewidth]{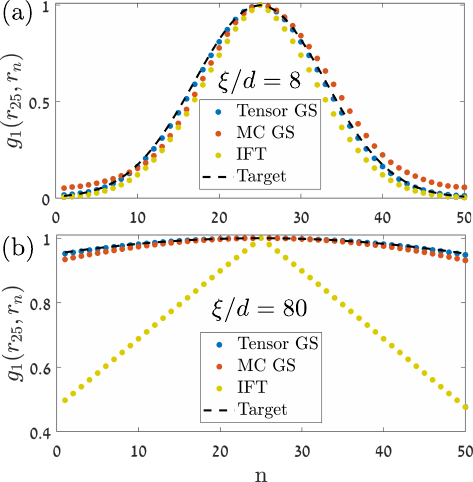}
    \caption{Reconstructed Gaussian-decaying mutual coherence function $g_1(r_{25}, r_{n})$ of a linear array of $N = 50$ Gaussian beams for two normalized coherence lengths (a) $\xi/d=8$ and (b) $ \xi/d=80$. Three reconstruction methods were used: Tensor GS (blue), MC GS using 100 realizations (red), and IFT (yellow). The target Gaussian-decaying mutual coherence function is denoted by the dashed curve (black).}
    \label{fig:2_linear_array}
\end{figure}

We simulated three methods for recovering the mutual intensity (Tensor GS, MC GS and IFT), using linear arrays of $N=50$ Gaussian beams in the near-field, with beam waist radii $w_0=100\mu m$, spacing between beams $d=300\mu m$, and with Gaussian-decaying spatial coherence of varying coherence lengths $\xi$. The results for two representative normalized coherence lengths, $\xi/d=8$ and $\xi/d=80$, are presented in Fig.~\ref{fig:2_linear_array}, along with the corresponding target mutual coherence. 

The spatial mutual coherence $g_1(r_{m}, r_{n})$, evaluated at the centers of the beams $m$ and $n$, where the reference beam $m=25$ is located at the center of the array. As evident, all  methods accurately reconstruct the mutual intensity function for $\xi/d = 8$.
However, for $\xi/d = 80$ where the coherence length exceeds the array size ($\xi > Nd$), the mutual intensity becomes strongly inhomogeneous. In this regime, the IFT method deteriorates significantly, whereas both the Tensor GS and MC GS methods remain accurate. Additional results demonstrating reconstruction of the full four-dimensional mutual intensity and off-center elements (e.g., m=10), are provided in Section 2 of Supplement.

\begin{figure}[ht!]
    \centering
    \includegraphics[width=0.65\linewidth]{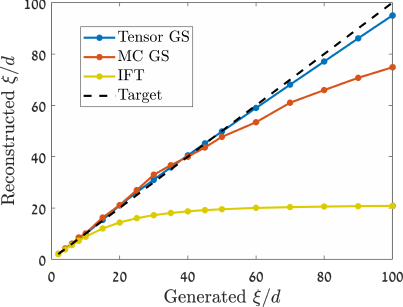}
    \caption{Reconstructed normalized coherence lengths $\xi/d$ as a function of the target values for a linear array of $N=50$ Gaussian beams. Three reconstruction methods were used: Tensor GS (blue), MC GS using 100 realizations (red), and IFT (yellow). The dashed line denotes the target coherence lengths.}
    \label{fig:3_linear_array_sum}
\end{figure}

For all three methods, the reconstructed normalized coherence lengths $\xi/d$ as a function of the target normalized coherence lengths are presented in Fig.~\ref{fig:3_linear_array_sum}. The reconstructed $\xi$ is obtained by fitting the mutual coherence to a Gaussian decay with the Gaussian width as a single fitting parameter. 
For $\xi \ll Nd$, the IFT method accurately reproduces the expected Gaussian coherence decay. However, as $\xi$ approaches the array size, finite-size effects in the far-field lead to significant deviations from the target coherence function. Yet, both Tensor GS and MC GS methods maintain accurate reconstruction of the spatial mutual intensity over a large range of $\xi$. As evident here and in Section 2 of Supplement, the Tensor GS method provides the most complete and accurate reconstruction of the full mutual intensity. Convergence rates of both methods are presented in Section 3 of Supplement. Results showing similar behavior were obtained for larger systems, including linear arrays of up to $N=600$ beams, see Section 4 of Supplement.

\begin{figure}[ht!]
    \centering
    \includegraphics[width=0.9\linewidth]{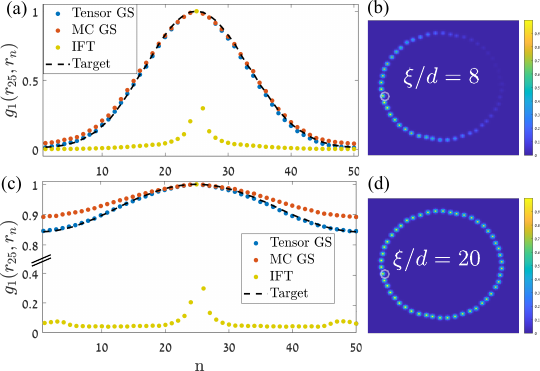}
    \caption{Reconstructed Gaussian-decaying mutual coherence function $g_1(r_{25},r_{n})$ of a ring array of $N = 50$ Gaussian beams for two normalized coherence lengths (a) $\xi/d=8$ and (c) $\xi/d=20$. Three reconstruction methods were used: Tensor GS (blue), MC GS using 1000 realizations (red), and IFT (yellow). The target Gaussian-decaying mutual coherence function is denoted by the dashed curves (black). (b) and (d) show the target mutual coherence in the near-field around the beam marked in a white circle. The mutual coherence decays as a Gaussian function that encircles the ring array and is summed up with itself, as indicated by the shape in the edges in (c).}
    \label{fig:4_Ring_Coherence_Detailed}
\end{figure}

\begin{figure}[ht!]
    \centering
    \includegraphics[width=0.68\linewidth]{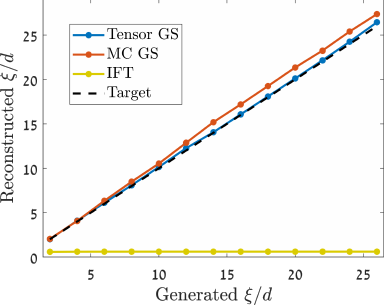}
    \caption{Reconstructed normalized coherence lengths $\xi/d$ as a function of the target values for a ring array of $N=50$ Gaussian beams. Three reconstruction methods are compared: Tensor GS (blue), MC GS using 1000 realizations (red), and IFT (yellow). The dashed line denotes the target coherence lengths.}
    \label{fig:5_Summary_ring_array}
\end{figure}

We performed a similar simulation for ring arrays of $N=50$ Gaussian beams in the near-field with beam waist radii $w_0=100\mu m$, spacing  $d=300\mu m$ and Gaussian-decaying spatial coherence along the ring of varying coherence lengths $\xi$. The results are presented in Figs.~\ref{fig:4_Ring_Coherence_Detailed} and~\ref{fig:5_Summary_ring_array}. For ring arrays the coherence function has rotational symmetry but it is strongly inhomogeneous even for small $\xi$. 

Figures~\ref{fig:4_Ring_Coherence_Detailed}(a) and~\ref{fig:4_Ring_Coherence_Detailed}(c) show the reconstructed mutual coherence $g_1(r_{m}, r_{n})$ around $m=25$ (representative of all sites up to rotational symmetry) obtained using the three reconstruction methods for two representative cases of $\xi/d=8$ and $\xi/d=20$. Even for moderate coherence lengths, the IFT method fails qualitatively due to the breakdown of the translational invariance assumption. In contrast, both the Tensor GS and MC GS methods accurately recover the correct structure of the mutual intensity. Notably, the Tensor GS method achieves high-fidelity reconstruction, while the MC GS method provides a computationally efficient approximation with good accuracy.  Figures~\ref{fig:4_Ring_Coherence_Detailed}(b) and~\ref{fig:4_Ring_Coherence_Detailed}(d) show the corresponding Gaussian-decaying coherence along the ring.

Figure~\ref{fig:5_Summary_ring_array} presents the reconstructed normalized coherence lengths $\xi/d$ as a function of the target values for the three reconstructions methods. The reconstructed $\xi$ is obtained by fitting the mutual coherence to a Gaussian function. The results obtained using the Tensor GS and MC GS methods demonstrate high reconstruction accuracy across the entire range, despite the strong spatial inhomogeneity of the coherence in the ring geometry. In contrast, the IFT method fails in this regime, consistent with the breakdown of its underlying assumptions.

\section{Experimental Results}

\begin{figure}[ht!]
    \centering
    \includegraphics[width=0.85\linewidth]{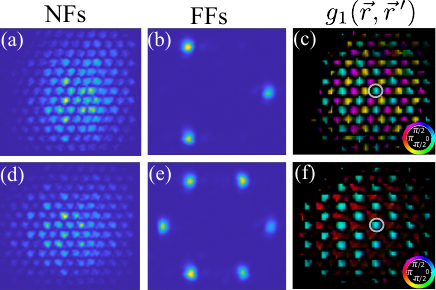}
    \caption{Experimental mutual coherence reconstruction using Tensor GS for a triangular array of 130 lasers with controlled coupling and staggered chirality. (a) Near-field and (b) far-field intensity distributions for the staggered anti-vortex state. (c) Reconstructed complex coherence for the staggered anti-vortex state around a reference laser marked in a white circle. (d) Near-field and (e) far-field intensity distributions for the equal superposition of staggered vortex and anti-vortex states. (f) Reconstructed complex coherence for the state of equal superposition around the same reference laser. In (c) and (f) color denotes phase and brightness denotes magnitude.}
    \label{fig:6_V_partial_coherence}
\end{figure}

To experimentally test our approach, we used a triangular array of 130 beams originating from 130 coupled lasers in a degenerate cavity with controlled coupling and staggered chirality, leading to states with non-monotonous mutual coherence decay and non-trivial phases, as elaborated in Section 5 of Supplement. We then reconstructed the mutual intensity distribution using our Tensor GS method, and the results are presented in Fig.~\ref{fig:6_V_partial_coherence}.

Figures~\ref{fig:6_V_partial_coherence}(a) and~\ref{fig:6_V_partial_coherence}(b) show the measured near-field and far-field intensity distributions for the staggered anti-vortex state. Figure~\ref{fig:6_V_partial_coherence}(c) shows the corresponding reconstructed complex mutual coherence function around a reference laser (marked by a white circle), reproducing the expected phase winding of $2\pi/3$ between neighboring lasers and a Gaussian decay of the mutual coherence magnitude. Figures~\ref{fig:6_V_partial_coherence}(d) and~\ref{fig:6_V_partial_coherence}(e) show the measured near-field and far-field intensity distributions for the superposition of the staggered vortex and the staggered anti-vortex states, and
Fig.~\ref{fig:6_V_partial_coherence}(f) shows the corresponding reconstructed complex mutual coherence function around the same reference laser. The reconstructed complex mutual coherence agrees well with that expected for the state of equal superposition which averages to $1$ where the vortex and anti-vortex states share the same phase (colored in cyan in Figs.~\ref{fig:6_V_partial_coherence}(c) and.~\ref{fig:6_V_partial_coherence}(f)), and to $-1/2$ otherwise. At the same time the spatial mutual coherence magnitude exhibits a Gaussian decay.

\begin{figure}[ht!]
    \centering
    \includegraphics[width=0.6\linewidth]{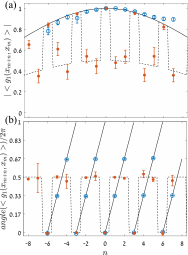}
    \caption{Quantitative comparison of experimentally measured and theoretically predicted cross-sections of reconstructed complex mutual coherence averaged over 16 lasers for the staggered anti-vortex state and the equal superposition of a staggered vortex and anti-vortex states. (a) Magnitude and (b) phase of the mutual coherence. Error bars indicate $80\%$ confidence interval. Blue circles: staggered anti-vortex state. Orange dots: the state of equal superposition. Solid black line: theoretical predictions for the staggered anti-vortex state. Dashed black line: theoretical predictions for superposition state.}
    \label{fig:7_VAV_partial_coherence}
\end{figure}

We also calculated and experimentally measured cross-sections of the mutual coherence and compared with theoretical predictions for both the staggered anti-vortex state and the equal superposition of staggered vortex and anti-vortex states. The results are presented in Fig.~\ref{fig:7_VAV_partial_coherence}, averaged over 16 reference lasers. 
Figure~\ref{fig:7_VAV_partial_coherence}(a) shows the magnitude and Fig.~\ref{fig:7_VAV_partial_coherence}(b) shows the complex phase of the mutual intensity. For the staggered anti-vortex state (blue circles), the magnitude exhibits a clear Gaussian decay, while the phase accumulates $2\pi/3$ between consecutive lasers, in agreement with theory (black solid line). 

For the state of equal superposition (orange dots), the theoretical mutual coherence exhibits a more complex structure, with alternating contributions corresponding to two Gaussian-decaying components (black dashed line): one with zero phase and full coherence, and a second with half coherence and an additional $\pi$ phase shift. This results in a spatially varying phase structure with an overall averaged phase that depends on the relative vortex–anti-vortex contributions.
The root-mean-square (RMS) phase error between experiment and theory is $2\pi/250$ for the staggered anti-vortex state and $2\pi/65$ for the state of equal superposition. The overall agreement demonstrates that the reconstruction method is valid across different types of partially coherent states and is robust to experimental noise.

\section{Conclusion}

In this work, we developed projection-based methods for reconstructing the mutual intensity and from it the phase distribution and mutual coherence of partially coherent optical fields from intensity measurements in two Fourier-conjugate planes. The Tensor GS method achieves high-accuracy reconstruction of the mutual intensity at the cost of substantial computational complexity, while the Monte Carlo (MC) GS method provides a computationally efficient controlled approximation. 

We validated both approaches on simulated inhomogeneous systems of coupled laser arrays with controlled coherence lengths in linear and ring geometries comprising up to $600$ beams. We then experimentally demonstrated the Tensor GS method on a triangular array of $130$ coupled lasers with controlled coupling and staggered chirality, reconstructing both the phase distribution and mutual coherence from measured near-field and far-field intensities with a root-mean-square phase error as low as $2\pi/250$, demonstrating robustness and practical applicability.

Our methods for coherence retrieval from intensity measurements, complement existing tomographic approaches, without requiring a predefined modal basis, low-rank parameterization, or multiple detection planes. They are best suited for mutual intensity functions whose spatial structure is linked to the geometry of the near-field intensity distribution. In particular, where the underlying inhomogeneous coherence is slowly varying relative to the near-field intensity.

Theoretically the uniqueness of mutual intensity reconstruction from intensity measurements in two Fourier-conjugate planes remains an open question. Nevertheless, our numerical and experimental results indicate stable convergence towards physically meaningful solutions for strongly inhomogeneous coherence structures and demonstrates robustness to experimental noise.

Future work can incorporate free-space propagation of the mutual intensity to develop phase and coherence reconstructions based on measured intensity in two or more non-Fourier-conjugate planes. This could result in more accurate coherence reconstructions without prior assumptions and for arbitrary optical fields and coherence functions. 
Moreover, the methods could be used to reconstruct statistical properties of the mutual intensity, such as coherent cluster size distributions in coupled laser systems. Finally, the framework developed here may be of interest to the quantum tomography community, given the formal analogy between mutual intensity reconstruction and density matrix reconstruction.

\begin{acknowledgments}
The authors thank the Minerva foundation for their support.
\end{acknowledgments}

\newpage
\addcontentsline{toc}{section}{References}
\bibliographystyle{ieeetr}
\bibliography{03_Bibliography.bib}

\newpage

\onecolumngrid
\appendix
\counterwithin{figure}{section}

\section{Monte Carlo Sequential Gerchberg Saxton}

We consider a simple and intuitive reconstruction method, termed MC Sequential GS. In this method, multiple independent coherent reconstructions are sequentially calculated using the standard GS that is based on the target intensities in the near-field and the far-field. The resulting mutual intensity is then calculated by treating the independent reconstructions as an ensemble, from which the spatial mutual coherence is obtained. We show that this approach serves as a useful baseline in cases where partial coherence arises from incoherent mixing of few fully coherent states. 

To understand the limits of MC Sequential GS, we simulated a ring array of $N=21$ Gaussian lasers with frustrated nearest-neighbor coupling producing a statistical mixture of two coherent vortex states with opposite topological charge. This arises from nearest-neighbor negative (out-of-phase) coupling, which for an odd number of lasers cannot produce an alternating $0,\pi$ phase configuration. Instead, such coupling results in one of the two states with topological charges $Q=\lfloor N/2 \rfloor=10$ and $Q=\lceil N/2 \rceil=11$ \cite{Pal_2015}. In this case, partial coherence arises not from stochastic decoherence, but from incoherent mixing of fully coherent fields. 

\begin{figure}[ht!]
    \centering
    \includegraphics[width=0.4\linewidth]{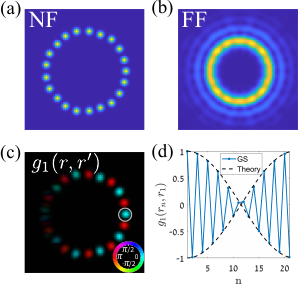}
    \caption{Reconstruction of coherence of an ensemble of equal vortex ($Q=10)$ and anti-vortex ($Q=11)$ states in a ring array with $N=21$ lasers. (a) and (b) Near-field and far-field average intensity distributions of the ensemble. (c) Reconstructed complex coherence around a reference Gaussian laser marked by white circle. Color denotes phase and brightness denotes magnitude. (d) Real part of the reconstructed complex coherence at each Gaussian center around the reference Gaussian laser $m=1$ compared to the theoretical values shown by the dashed curve.}
    \label{fig:A1_odd_OOP}
\end{figure}

The results of the simulations are presented in Fig.~\ref{fig:A1_odd_OOP}. Figures~\ref{fig:A1_odd_OOP}(a) and~\ref{fig:A1_odd_OOP}(b) show the ensemble-averaged near-field and far-field intensity distributions for the ensemble. Figure~\ref{fig:A1_odd_OOP}(c) shows the reconstructed two-dimensional complex coherence $g_1(r_{n},r_{m})$, of MC Sequential GS with 100 realizations, around laser number $m=1$ marked in a white circle (coherence around a different laser is achieved by rotational symmetry). Figure~\ref{fig:A1_odd_OOP}(d) shows the reconstructed coherence at the center of every laser. As evident, it matches well the theoretically predicted envelope of the coherence decay. Surprisingly, the MC Sequential GS succeeds in reconstructing the complex coherence function, and we found that it can do so even when limiting the ensemble size to $N=2$ realizations.

\begin{figure}[ht!]
    \centering
    \includegraphics[width=0.45\linewidth]{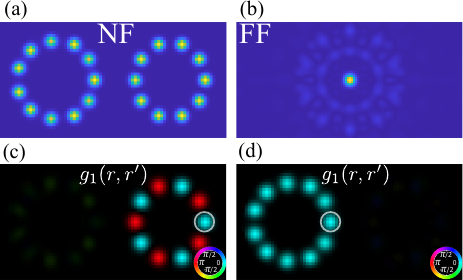}
    \caption{Reconstruction of partial coherence of an ensemble of fields in side-by-side ring arrays of lasers that are fully coherent within each array, one in-phase and one out-of-phase, and incoherent with each other. (a) and (b) Average near-field and far-field intensity distributions of the ensemble. Reconstructed complex coherence around (c) one laser on the right array, with $N=10$ lasers, and (d) one laser on the left array, with $N=11$ lasers, both marked by white circles. Reconstruction by MC Sequential GS using 100 realizations. Color denotes phase and brightness denotes magnitude.}
    \label{fig:A2_EvenOdd}
\end{figure}

We also examined two side-by-side ring arrays of $N=10$ and $N=11$ Gaussian lasers with equal spacing that are coherent within and mutually incoherent. We simulated the ensemble-averaged near-field and far-field intensity distributions for such an ensemble, shown in Figs.~\ref{fig:A2_EvenOdd}(a) and~\ref{fig:A2_EvenOdd}(b), and from them reconstructed the complex coherence using MC Sequential GS with 100 realizations. Figures~\ref{fig:A2_EvenOdd}(c) and~\ref{fig:A2_EvenOdd}(d) present the reconstructed two-dimensional complex mutual coherence $g_1(r,r')$ when centered around a laser in the right ring array and the left ring array respectively, both marked in white circles. MC Sequential GS successfully recovers intra-array coherence and inter-array incoherence when the spatial mutual coherence is correlated to the array geometry. It, serves as a useful baseline in cases where partial coherence arises from incoherent mixing of few fully coherent states.

\section{Exploration of the Four-Dimensional Mutual Intensity}

\begin{figure}[ht!]
    \centering
    \includegraphics[width=0.7\linewidth]{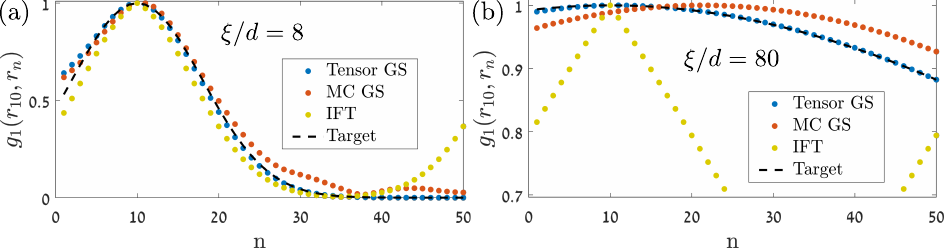}
    \caption{Reconstructed Gaussian-decaying mutual coherence function $g_1(r_{10}, r_{n})$ of a linear array of $N = 50$ Gaussian beams for two normalized coherence lengths (a) $\xi/d=8$ and (b) $ \xi/d=80$. Three methods were used: Tensor GS (blue), MC GS using 100 realizations (red), and IFT (yellow). The target Gaussian-decaying coherence function is denoted by the dashed curve (black).}
    \label{fig:B1_linear_array_side}
\end{figure}

To further explore the four-dimensional reconstruction of the mutual intensity and mutual coherence, we plot different cuts of the mutual coherence for the linear arrays of $N=50$ Gaussian beams in the near-field with Gaussian-decaying coherence of varying coherence lengths $\xi$. The results for representative normalized coherence lengths, $\xi/d=8$ and $\xi/d=80$ are presented in Fig.~\ref{fig:B1_linear_array_side}. These are the same normalized coherence lengths presented in Fig.~2, but in Fig.~\ref{fig:B1_linear_array_side} the spatial mutual coherence $g_1(r_{m}, r_{n})$ is for $m=10$ (not $m=25$), a side beam in the array. The spatial inhomogeneity of the mutual coherence can be observed by comparing the target mutual coherence in Fig.~2 and Fig.~\ref{fig:B1_linear_array_side}. As evident, also for $m=10$ the Tensor GS and MC GS methods accurately reconstruct the mutual coherence function. For $\xi/d = 80$ the Tensor GS method reconstructs the mutual coherence to high accuracy indicating the strength and robustness of the method in reconstructing strongly inhomogeneous mutual coherence functions.

\newpage

\section{Monte Carlo Gerchberg Saxon Method Convergence}

We quantify the convergence of the Tensor GS and MC GS methods when using linear arrays of $N=50$ Gaussian beams in the near-field with Gaussian-decaying coherence of varying coherence lengths $\xi$. We calculate the normalized root-mean-square difference between the target intensities in the near-field and far-field and the target mutual intensity compared to the ensemble averaged values for different iterations of the algorithm. The normalization is performed by dividing by the maximal value of the target function to get unitless values. For the calculation of the near-field and far-field intensities and mutual intensity we use Eq. (\ref{EqC1_RMS_convergence}). The mutual intensity function is calculated for a specific $\mathbf{r'}$ chosen at the center of the central beam in the array. Note that $<\cdot>_\mathbf{r}$ denotes averaging over the spatial degrees of freedom while $<\cdot>_{MC}$ denotes averaging over the realizations of the Monte Carlo ensemble.

\begin{equation}
    \begin{split}
        & \Delta I_{NF} = \frac{1}{\max_\mathbf{r}\{I_{\text{NF Target}}(\mathbf{r})\}} \sqrt{<|I_{\text{NF Target}}(\mathbf{r})-<|E(\mathbf{r})|^2>_{\text{MC}}|^2>_\mathbf{r}} \hspace{2mm},\\
        & \Delta I_{FF} = \frac{1}{\max_\mathbf{r}\{I_{\text{FF Target}}(\mathbf{k})\}} \sqrt{<|I_{\text{FF Target}}(\mathbf{k})-<|FT\{E\}(\mathbf{k})|^2>_{\text{MC}}|^2>_\mathbf{k}} \hspace{2mm},\\
        & \Delta J_{NF} = \frac{1}{\max_\mathbf{r}\{J_{\text{NF Target}}(\mathbf{r},\mathbf{r'})\}} \sqrt{<|J_{\text{NF Target}}(\mathbf{r},\mathbf{r'})-<E^* (\mathbf{r})E(\mathbf{r'})>_{\text{MC}}|^2>_\mathbf{r}} \hspace{2mm}.
    \end{split}
    \label{EqC1_RMS_convergence}
\end{equation}

\begin{figure}[ht!]
    \centering
    \includegraphics[width=0.9\linewidth]{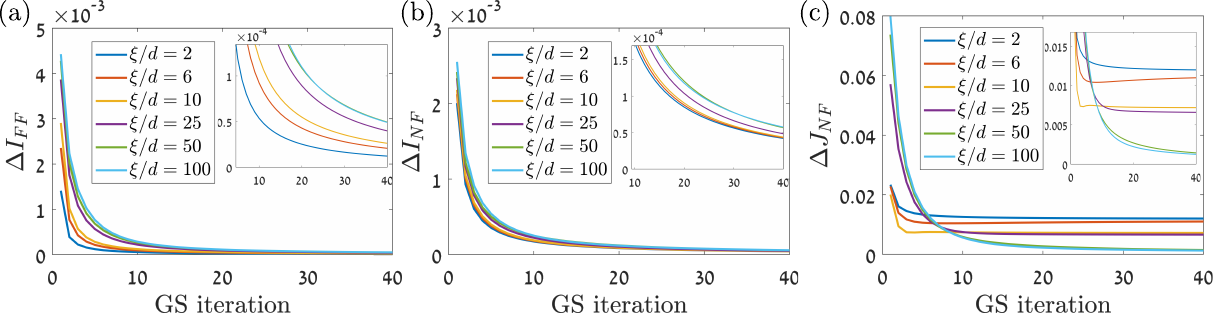}
    \caption{Convergence of the near- and far-field intensities and the mutual intensity to the target values as a function of iteration number using MC GS. The calculated root-mean-square distance from the target values for (a) near-field, (b) far-field and (c) mutual intensity, in a linear array of $N = 50$ Gaussian beams, and for 6 normalized coherence lengths $\xi/d=2,6,10,25,50,100$. Insets show the same results when zooming in at the end of the convergence}
    \label{fig:C1_MC_GS_Converge}
\end{figure}

The results for MC GS for 40 iterations are presented in Fig.~\ref{fig:C1_MC_GS_Converge} for 6 representative normalized coherence lengths $\xi/d$. As evident, for all normalized coherence lengths the intensities in the near- and far-field, and the mutual intensity converge to the target values as the root-mean-square distance approaches zero in less than 40 iterations.

\begin{figure}[ht!]
    \centering
    \includegraphics[width=0.9\linewidth]{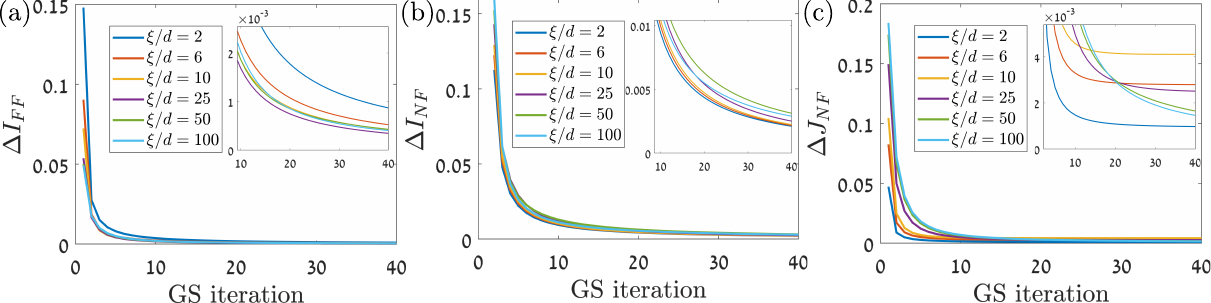}
    \caption{Convergence of the near- and far-field intensities and the mutual intensity to the target values as a function of iteration number using Tensor GS. The root-mean-square distance from the target values is calculated for the (a) near-field, (b) far-field and (c) mutual intensity for 40 iterations of Tensor GS for a linear array of $N = 50$ Gaussian beams for 6 normalized coherence lengths $\xi/d=2,6,10,25,50,100$. Insets show the results zoomed in at near convergence}
    \label{fig:C1_Tensor_GS_Converge}
\end{figure}

The Tensor GS results for 40 iterations are presented in Fig.~\ref{fig:C1_Tensor_GS_Converge} for 6 representative normalized coherence lengths $\xi/d$. As evident, for all normalized coherence lengths the intensities in the near- and far-fields, and the mutual intensity converge to the target values as the root-mean-square distance approaches zero in less than 40 iterations.

\section{Reconstruction of the Mutual Intensity for 600 Beams}

\begin{figure}[ht!]
    \centering
    \includegraphics[width=0.7\linewidth]{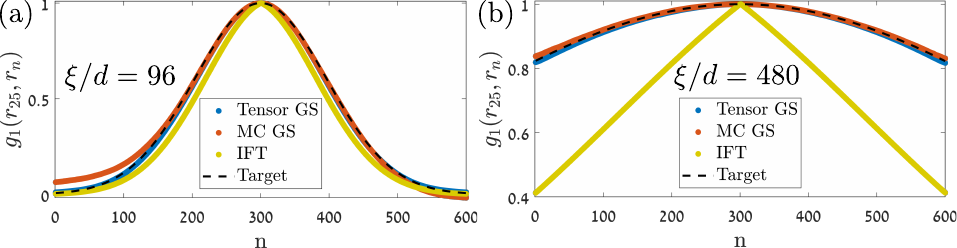}
    \caption{Reconstructed Gaussian-decaying mutual coherence function $g_1(r_{300}, r_{n})$ of a linear array of $N = 50$ Gaussian beams for two normalized coherence lengths (a) $\xi/d=96$ and (b) $ \xi/d=480$. Three methods were used: Tensor GS (blue), MC GS using 1000 realizations (red), and IFT (yellow). The target Gaussian-decaying coherence function is denoted by the dashed curve (black).}
    \label{fig:D1_linear_array_N600}
\end{figure}

We demonstrate mutual intensity reconstruction with Tensor GS and MC GS for the linear arrays of $N=600$ Gaussian beams in the near-field with Gaussian-decaying coherence of varying coherence lengths $\xi$. The results for representative normalized coherence lengths, $\xi/d=96$ and $\xi/d=480$ are presented in Fig.~\ref{fig:D1_linear_array_N600}. These are the same normalized coherence lengths presented in Fig.~2, but Fig.~\ref{fig:D1_linear_array_N600} shows the spatial mutual coherence $g_1(r_{m}, r_{n})$ for $m=300$ the central beam in the array. As evident, also for $N=600$ the Tensor GS and MC GS methods accurately reconstruct the mutual coherence function. 

\section{Staggered Vortex and Anti-Vortex States in Coupled Lasers}

In triangular arrays of coupled lasers, the phase configuration of the lowest-loss lasing state is determined by the geometry and sign of the nearest-neighbor coupling. For negative nearest-neighbor coupling, two neighboring lasers will phase-lock in an out-of-phase configuration. However, in a triangular lattice this condition cannot be simultaneously satisfied for all pairs of nearest-neighbors, resulting in geometric frustration. Consequently, the lowest-loss states exhibit nontrivial phase winding corresponding to staggered vortex and anti-vortex configurations with alternating chirality across the array.

\begin{figure}[ht!]
    \centering
    \includegraphics[width=0.35\linewidth]{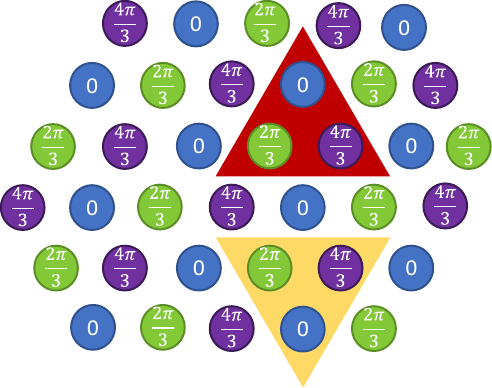}
    \caption{Coupled laser in a triangular array in a staggered vortex state colored by the complex phases of the lasers. A triangle of nearest-neighbor lasers with positive chirality is marked in red and one with negative chirality is marked in yellow.}
    \label{fig:E1_V_AV_flux}
\end{figure}

Figure~\ref{fig:E1_V_AV_flux} illustrates a staggered vortex state in a triangular array. Neighboring triangular pads possess opposite chirality, as highlighted by the red and yellow triangles. The phase accumulated around each pad corresponds to a discrete vortex flux.

To experimentally control these states, we resort to a mid-field coupling scheme \cite{Mahler_2025}, which enables complex coupling between neighboring lasers and effectively realizes an artificial gauge field (AGF). The resulting coupling phase can be interpreted as an effective magnetic flux through each triangular plaquette of the array.

Positive flux favors staggered anti-vortex states, while negative flux favors staggered vortex states. When the flux approaches zero, the two states become nearly degenerate and coexist in different longitudinal cavity modes, leading to partially coherent superpositions of staggered vortex and anti-vortex configurations.

These states provide an experimentally relevant example of partially coherent fields with both nontrivial phase structure and spatially varying mutual coherence. In particular, the coexistence of vortex and anti-vortex states generates coherence functions with nontrivial spatial phase distributions and Gaussian-decaying coherence magnitude, making them a strong test case for the proposed coherence reconstruction methods.

\end{document}